\documentclass[10pt,conference]{IEEEtran}

\newcommand{\LIF}[2]{ \STATE \algorithmicif\ {#1}\ \algorithmicthen\ {#2}}
\newcommand{\LIFELS}[3]{ \STATE \algorithmicif\ {#1}\ \algorithmicthen\ {#2}\ \algorithmicelse\ {#3}}

\newcommand{\LFOR}[2]{\STATE\algorithmicfor\ {#1}\ \algorithmicdo\ {#2}}

\newtheorem{theorem}{Theorem}
\newtheorem{lemma}[theorem]{Lemma}

\ifCLASSINFOpdf
   \usepackage[pdftex]{graphicx}

  \DeclareGraphicsExtensions{.pdf,.jpeg,.png}
\else
   \usepackage[dvips]{graphicx}
   \graphicspath{{../eps/}}
   \DeclareGraphicsExtensions{.eps}
\fi

\usepackage{amsmath}
\ifCLASSOPTIONcompsoc
  \usepackage[nocompress]{cite}
\else
  \usepackage{cite}
\fi

\usepackage{algorithmic}

\usepackage{epsfig,endnotes}
\usepackage{url}
\usepackage{multirow}
\usepackage{stfloats}
\usepackage{amssymb}
\usepackage[export]{adjustbox}

\usepackage{algorithm}
\usepackage{multirow}

\title{
Making  Name-Based Content Routing More Efficient than Link-State Routing
}

\author{Ehsan Hemmati$^{*}$ and 
J.J. Garcia-Luna-Aceves$^{*, \dagger}$ \\
$^{*}$Computer Engineering Department, 
UC Santa Cruz, Santa Cruz, CA 95064 \\
$^{\dagger}$PARC, Palo Alto, CA  94304\\
\{ ehsan, jj \}@soe.ucsc.edu
}

\IEEEoverridecommandlockouts

\IEEEpubid{
\begin{minipage}{\textwidth}\ \\[12pt]
\copyright~IFIP, 2018 This is the author's version of the work. It is posted here by\hfill\\permission of IFIP for your personal use. Not for redistribution. {ISBN 978-3-903176-08-9, (2018)}  \hfill
\end{minipage}
}

\begin{document}

\maketitle

\begin{abstract}

The Diffusive Name-based Routing  Protocol (DNRP)  is  introduced for efficient  name-based  routing  in  information-centric networks (ICN).   DNRP establishes and maintains multiple loop-free routes to the nearest instances of a name prefix using  only distance information.  DNRP eliminates the need for periodic updates, maintaining topology information, storing complete paths to content replicas, or knowing about all the sites storing replicas of named content.  DNRP is suitable for large ICNs with large numbers of prefixes stored at multiple sites. It is shown that DNRP  provides loop-free routes to content independently of the state of the topology, and that it converges within a finite time to correct routes to name prefixes after arbitrary changes in the network topology or the placement of  prefix instances.  The result of simulation experiments illustrate  that DNRP is more efficient than  link-state routing approaches.
\end{abstract}

\section{Introduction}

Several Information-Centric Networking (ICN) architectures have been introduced to  address the increasing demand of user-generated content \cite{Bari2012, Choi2011}. The goal of these architectures is to provide a cost-efficient, scalable, and mobile access to content and services by adopting a content-based model of communication. ICN architectures seek to dissociate content and services  from their producers in such  a way that the content can be retrieved independently of its original location or the location of consumers. 
The most prominent ICN architectures can be characterized as Interest-based architectures,  in which location-independent, self-defined, unique naming is used to retrieve data.  In this approach, messages flow from producers to consumers based on the name of the content  rather than the address of the senders or receivers exchanging such content. Content providers or producers create named data objects (NDOs), and advertise routable name prefixes associated with the content objects whose own names are part of the name prefixes. The  only identifier of an NDO is its name. A consumer requests a piece of content by sending an Interest  (a request for the NDO) that is routed along content routers toward the producer(s). 

Clearly, an efficient  name-based content routing protocol must  be used for any ICN architecture to succeed using name-based forwarding of Interests and  requested content.
Section \ref{sec-related} summarizes recent prior work in name-based content routing, and this review reveals that all prior proposals for name-based content routing rely on periodic transmissions of update messages.
This paper focuses on an approach that avoids the need for periodic messaging by means of 
diffusing computations \cite{dual}.

Section \ref{sec-DNRP} presents {\bf DNRP} ({\em Diffusive Name-based Routing Protocol}), a name-based content routing protocol for ICNs. 
DNRP provides multiple loop-free routes to the nearest instances of a named prefix or to all instances of a named prefix  using only distance information and without requiring  periodic updates, knowledge of the network topology,  or the exchange of path information. 

Section \ref{sec-correct} shows that DNRP  prevents routing-table loops even in the presence of topology changes as well as changes in the hosting of  prefixes, and converges within a finite time to correct multi-paths to name prefixes.  Section \ref{sec-sim} presents the results of simulation experiments comparing DNRP with an efficient  link-state approach similar to NLSR  \cite{nlsr2}. The results show that DNRP produces less communication and computation overhead in the case of topology changes as well as the addition of prefixes.

\section{Related Work}
\label{sec-related}
	


Name-based content routing has been used in the past in content-delivery networks (CDN) operating as  overlay networks running on top of the Internet routing infrastructure (e.g.,  \cite{wild, gold}). However, it has become more well known in the context of  ICN architectures, where it has been done typically 
by adapting traditional routing algorithms designed for networks in which a destination has a single instance \cite{garcia2014name}.  Distributed Hash Tables (DHT) are used in several architectures as the name resolution tool \cite{PURSUIT,sail,Katsaros2012}. MobilityFirst \cite{mob} rely on an external and fast name resolution system called Global Name Resolution Service (GNRS) that maps the data object names to network addresses. 
	
	Some ICN architectures use name resolution mechanisms to map the name of the content to the content provider.  Data Oriented Network Architecture (DONA) \cite{Koponen2007} replaces DNS names with ``flat, self-certifying
names"  and uses name resolution to map those flat names to corresponding IP addresses. DONA supports  host mobility and multihoming, and improves service access and data retrieval.
	
		The Named-data Link State Routing protocol (NLSR) \cite{nlsr2} 
uses link state routing to rank the neighbors of a router for each name prefix. "€œAdjacency LSA"€ and "€œPrefix LSA"€, propagate topology and publisher information in the network respectively. 
Each router uses topology information and runs an extension of Dijkstra's shortest-path first (SPF) algorithm to rank next hops for each router, then maps the prefix to the name of the publisher and creates routing table for each name prefix. 
Like most prior routing approaches based on complete or partial  topology information (e.g.,  \cite{hlvr, icnp98, nasr}), NLSR uses sequence numbers to allow routers to determine whether the updates they receive have more recent information than the data they currently store. As a consequence,   these approaches require the use of periodic updates to percolate throughout the network
to ensure that all routers converge to consistent topology information within a finite time. 

	The Distance-based Content Routing (DCR) \cite{garcia2015fault} was the first name-based content routing approach for ICNs based on distances to named content. DCR does not require any information about the network topology or knowledge about all the instances of the same content. It enables routing to the nearest router announcing content from a name prefix being  stored locally (called anchor), all anchors of a name prefix, and subsets of anchors.
This is attained by means of  multi-instantiated destination spanning trees (MIDST).  Furthermore, 
DCR provides loop-free routes to reach any piece of named content even if different content objects in the same prefix are stored at different sites. The limitation of DCR is that  it requires periodic updates to be disseminated through the network.

\section{DNRP}
\label{sec-DNRP}

	DNRP finds the shortest path(s)  to the nearest replica(s) of name prefixes.  To ensure that loop-free routes to named prefixes are maintained at every instant independently of the state of the network or prefixes, DNRP establishes a lexicographic ordering among the routes to prefixes  reported and maintained by routers. The lexicographic ordering of routes is based on two sufficient conditions for loop freedom with respect to a given prefix that allow for multiple next hops to prefixes along loop-free routes. DNRP diffuses  the computation of new loop-free routes when the loop-free conditions are not satisfied.
The  approach used in DNRP  is an extrapolation to the use of diffusing computations in  \cite{dual}. 

Every piece of data  in the network is a {\it Named-Data Object} ({\em NDO}), represented by a name that 
belongs to a name prefix or simply a {\it prefix} advertised by one or more producer(s). 
Name prefixes can be simple and human-readable or more complicated and self certifying, or may even be a cryptographic hash of the content. Content names can be flat or hierarchical. The naming schema depends on the system that runs DNRP and it is out of scope of this paper.  

A router attached to a producer of content that advertises a name prefix is called an  $anchor$ of that prefix.  At each router, DNRP calculates routes to the nearest anchor(s) of known name prefixes, if there is any, and selects a subset of the neighbors of the router as valid next hops to reach  name prefixes, such that no routing-table loop is created at any router for any name prefix. Caching sites are not considered content producers and hence routes to cached content are not advertised in DNRP.  Our description assumes that routers process, store, and transfer information correctly and that they process routing messages one at a time within a finite time. Every router  has a unique identifier or a name that can be flat or hierarchical. The naming schema and name assignment mechanism is out of scope of this paper.

\subsection{Messages and Data Structures}
	
Each router $i$ stores the list of all active neighbor routers ($N^i$), and the cost of the link from the router to each such neighbor. The cost of the link from router $i$ to its neighbor $n$  is denoted by $l^i_n$. Link costs can vary in time but  are always positive. The link cost assignment and metric determination mechanisms are beyond the scope of this paper.
	
	The routing information reported by each of the neighbors of router $i$  is stored in its {\it neighbor table} ($NT^i$). The entry of $NT^i$ regarding neighbor $n$ for prefix $p$ is denoted by $NT^i_{pn}$ and consists of the name prefix ($p$), the distance to prefix $p$ reported by neighbor $n$ ($d^i_{pn}$), and the anchor of that prefix reported by neighbor $n$ ($a^i_{pn}$). If router $i$ is the anchor of prefix $p$ itself, then $d^i_{pi}$ = 0.
	
	Router $i$ stores routing information for each known prefix in its routing table ($RT^i$). The entry in $RT^i$ for prefix $p$ ($RT^i_p$) specifies: the name of the prefix ($p$); 
the distance to the nearest instance of that prefix ($d^i_p$); the feasible distance to the prefix ($fd^i_p$); the neighbor that offers the shortest distance to the prefix 
($s^i_p$), which we call the successor of the prefix; the closest anchor to the prefix ($a^i_p$); the state mode ($mf^i_p$) regarding prefix $p$, which can be {\it PASSIVE} or {\it ACTIVE}; the origin state ($o^i_p$) indicating whether router $i$ or a neighbor is the origin of query in which  the router is active; the update flag list ($FL^i_p$); and the list of all valid next hops ($V^i_p$).  
	
$FL^i_p$ consists of four flags for each neighbor $n$. An update flag      ($uf^i_{pn}$)  denotes whether or not the routing message should be sent to that neighbor. A type flag ($tf^i_{pn}$) indicates the type of routing message the router has to send to neighbor $n$ regarding prefix $p$ (i.e., whether  it is an {\it UPDATE}, {\it QUERY}, or {\it REPLY}). A pending-reply flag ($rf^i_{pn}$) denotes whether the router has sent a {\it QUERY} to that neighbor and is waiting for {\it REPLY}. A pending-query flag ($qf^i_{pn}$)  is set if the router received a {\it QUERY} from its neighbor $n$ and has not responded to that {\it QUERY} yet. 
		
	Router $i$ sends a routing message to each of its neighbors containing updates made to $RT^i$ since the time it sent its last
update message. A routing message from router $i$ to neighbor $n$ consists of one or more updates, each of which carries information regarding one prefix that needs updating. The update information for prefix $p$ is denoted by $U^i_p$ and states: (a) the name of the prefix ($p$); (b) the message type ($ut^i_p$) that indicates if the message is an {\it UPDATE}, {\it QUERY}, or {\it REPLY}; (c) the distance to $p$; and (d) the name of the closest anchor.	
	
	The routing update received by router $i$ from neighbor $n$ is denoted by $U^i_n$. The update information of $U^i_n$ for prefix $p$, $u^i_{pn}$,  specifies the prefix name ($p$), the distance from $n$ to that prefix ($ud^i_{pn}$), the name of the anchor of that prefix ($ua^i_{pn}$), and a message type ($ut^i_{pn}$).

\subsection{Sufficient Conditions for Loop Freedom }

The 
conditions for loop-free routing in DNRP are based on  the feasible distance maintained at each router and the distances reported by its neighbors for a name prefix. One condition is used to determine the new shortest distance through a loop-free path to a name prefix. The other is used to select a subset of neighbors as next hops to a name prefix.

	

\vspace{0.05in}	
{\bf Source Router Condition (SRC):}
Router $i$ can select  neighbor $n \in N^i$ as a new successor $s^i_p$ for prefix $p$ if:
\[
\begin{split}
(~ d^i_{pn}  <  \infty ~)  ~\wedge
(~ d^i_{pn}  <  fd^i_{p}  \vee
[d^i_{pn}  = fd^i_{p} \wedge |n| < |i|] ~) ~\wedge \\
(~ d^i_{pn} + l^i_n = Min \{d^i_{pv} + l^i_v | v \in N^i \} ~). ~~ \Box
\end{split}
\]

{\it SRC} simply states that router $i$ can select neighbor $n$ as its successor to prefix $p$ if $n$ reports a finite distance to that prefix, offers the smallest distance to prefix $p$ among all neighbors, and either its distance to prefix $p$ is less than the feasible distance of router $i$ or its distance is equal to the feasible distance of $i$ but $|n| < |i|$. If two or more neighbors satisfy  {\it SRC}, the neighbor that satisfies {\it SRC} and has the smallest identifier is selected. If none of the neighbors satisfies  {\it SRC} the  router keeps the current successor, if it has any. The distance of router $i$ to prefix, $d^i_p$, is defined by the distance of the path through the selected successor. 
	
	A router that has a finite feasible distance ($fd^i_p < \infty$) selects a subset of neighbors as valid next hops at time $t$ if they have a finite distance to destination and are closer to destination. The following condition is used for this purpose.

\vspace{0.05in}	
{\bf Next-hop Selection Condition (NSC):}
Router $i$ with $fd^i_p < \infty $ adds neighbor $n \in N^i$  to the set of valid next hops if:
$(~ d^i_{pn}  <  \infty ~)  \wedge 
(~ d^i_{pn}  <  fd^i_{p} \vee 
[d^i_{pn}  =  fd^i_{p} \wedge |n| < |i|]~ )$. $\Box$


\vspace{0.05in}		
{\it NSC} states that router $i$ can select neighbor $n$ as a  next hop to prefix $p$ if either the  distance from $n$ to prefix $p$ is smaller than the feasible distance of $i$ or its distance is equal to the feasible distance and $|n| < |i|$. {\it NSC} orders next hops lexicographically based on their distance to a prefix and their names. It is shown that no routing-table loops can be formed if {\it NSC} is used to select the next hops to prefixes at each router. Note that the successor is also a valid next hop. The  successor to a prefix is a valid next hop that offers the smallest cost.

{\it SRC} and {\it NSC} are  {\it sufficient conditions} that, as we show subsequently,  ensure loop-freedom at every instance but  do not guarantee shortest paths to destinations. DNRP integrates  these sufficient conditions with 
inter-nodal synchronization signaling  to achieve both loop freedom at every instant and shortest paths for each destination. 

\subsection{DNRP Operation}
	
	A change in the network, such as a link-cost change, the addition or failure of a link, the addition or failure of a router, the addition or deletion of a prefix, or the addition or deletion of a replica of a prefix can cause one or more computations at each router for one or more prefixes. A computation can be either a {\it local computation} or a {\it diffusing computation}. In a local computation a router updates its successor, distance, next hops, and feasible distance independently of other routers in the network. On the other hand, in a diffusing computation a router originating the computation must coordinate  with other routers before making any changes in its 
routing-table entry for a given prefix.  DNRP allows a  router to participate in at most one diffusing computation per prefix at any given time. 
	
	A router can be in {\it PASSIVE} or {\it ACTIVE} mode with respect to a given prefix independently of other prefixes. A router is {\it PASSIVE} with  respect to prefix $p$ if it is not engaged in any diffusing computation regarding that prefix. A router initializes itself in {\it PASSIVE} mode and with a zero distance to all the prefixes for which the router itself serves as an anchor. An infinite distance is assumed to any non-local (and hence unknown) name prefix.
	
	Initially, no router is engaged in a diffusing computation ($o^i_p = 0$). When a {\it PASSIVE} router detects a change in a link or receives a {\it QUERY} or {\it UPDATE} from its neighbor that does not affect the current successor or can find a feasible successor, it remains in {\it PASSIVE} mode. On the other hand, if the router cannot find a feasible successor then it enters the {\it ACTIVE} mode and keeps the current successor, updates its distance, and sends {\it QUERY} to all its neighbors.
	 Table \ref{tbl:statetransit} shows the transit from one state to another. Neighbor $k$ is a neighbor other than the successor $s$.

\vspace{-0.1in}
\begin{table}[h]
\renewcommand{\arraystretch}{1.3}
\centering
\caption{State transit in DNRP}
\label{tbl:statetransit}
{\fontsize{7}{8}\selectfont
\begin{tabular}{|l|l|l|l|}
\hline
\textit{Mode}                     & \textit{State}     & Event                                      & \textit{\begin{tabular}[c]{@{}l@{}}Next \\ State\end{tabular}} \\ \hline
\multirow{3}{*}{\textit{Passive}} & \multirow{3}{*}{0} & Events from a neighbor $k$, {\it SRC} satisfied    & 0                                                              \\ \cline{3-4} 
                                  &                    & Events from a neighbor $k$, {\it SRC} not satisfied & 1                                                              \\ \cline{3-4} 
                                  &                    & {\it QUERY} from the {\it Successor}                       & 3                                                              \\ \hline
\multirow{10}{*}{\textit{Active}}          & \multirow{3}{*}{1} & Receives last {\it REPLY}                        & 0                                                              \\ \cline{3-4} 
                                  &                    & Change in distance to {\it Successor}             & 2                                                              \\ \cline{3-4} 
                                  &                    & {\it QUERY} from the {\it Successor}                        & 4                                                              \\ \cline{2-4} 
                                  & \multirow{3}{*}{2} & Receives last {\it REPLY}, {\it SRC} satisfied         & 0                                                              \\ \cline{3-4} 
                                  &                    & Receives last {\it REPLY}, {\it SRC} not satisfied     & 1                                                              \\ \cline{3-4} 
                                  &                    &  {\it QUERY} from the {\it Successor}                       & 4                                                              \\ \cline{2-4} 
                                  & \multirow{2}{*}{3} & Receives last {\it REPLY}                        & 0                                                              \\ \cline{3-4} 
                                  &                    & Change in distance to the {\it Successor}             & 4                                                              \\ \cline{2-4} 
                                  & \multirow{2}{*}{4} & Receives last {\it REPLY}, {\it SRC} satisfied   & 0                                                              \\ \cline{3-4} 
                                  &                    & Receives last {\it REPLY}, {\it SRC} not satisfied     & 3                                                              \\ \hline
\end{tabular}}
\end{table}

	Algorithm~\ref{alg:passive}  shows the  processing of messages by a router  in {\it PASSIVE} mode. Algorithm~\ref{alg:o0} shows the steps taken in {\it ACTIVE} mode.  Algorithm~\ref {alg:updateRTinf}  shows the steps taken to process a routing update.
	
\vspace{0.05in}			
	{\bf Handling A Single Diffusing Computation:} 
Routers are initialized in {\it PASSIVE} mode. Each router continuously monitors its links and processes  the routing messages received from its neighbors. When router $i$ detects a change in the cost or state of a link, or a change in its neighbor table that causes a change in its distance to prefix $p$ ($d^i_p$), it first tries to select a new successor that satisfies {\it SRC}. If such a successor exists, the router carries out a local computation, updates its distance, successor, and closest anchor, and exits the computation. In a local computation, router $i$ computes the minimum cost to reach the destination and updates $d^i_p = min\{d^i_{pn}+l^i_n | n \in N^i\}$. If its  distance changes, router $i$ sends a routing message with $ut^i_p = {\it UPDATE}$. Router $i$ also updates its feasible distance to equal the smaller  of its  value and the new distance value, i.e., $fd^i_p(new) = min \{fd^i_p(old), d^i_p\}$.
	
	An {\it UPDATE} message from a neighbor is processed using the same approach stated above. If a router receives a {\it QUERY} from its neighbor other than its successor while it is in {\it PASSIVE} mode, it updates the neighbor table, checks for a feasible successor according to {\it SRC} and replies with $d^i_p$, if it succeeds.
If router $i$ cannot find a neighbor that satisfies {\it SRC} after a change in a link or neighbor-table entry, then it starts out a diffusing computation by setting the new distance as the distance through its current successor, enters the {\it ACTIVE}  mode ($mf^i_j =$ {\it ACTIVE}) and sets the corresponding flag ($rf^i_pn$) for each neighbor $n$. After entering the {\it ACTIVE} mode, router $i$ sets the new distance as the cost of the path through the current successor ($d^i_p = d^i_{ps^i_p}+l^i_{s^i_p}$) and sends a routing message with $ut^i_p =$ {\it QUERY}. Router $i$ uses the pending reply flag ($rf^i_{pn}$) to keep track of the neighbors  from which a {\it REPLY} has not been received. When a router becomes {\it ACTIVE} it sets the update flag ($uf^i_{pn} = 1$), and also sets the type flags ($tf^i_{pn} = QUERY| \forall n \in N^i$) and sends the routing messages to all its neighbors.

\vspace{-0.15in}
\begin{figure}[h]
\begin{algorithm}[H]
\caption{Processing routing messages in {\it PASSIVE} mode}
\label{alg:passive}
{\fontsize{9}{9}\selectfont
\begin{algorithmic}
    \STATE{\textbf{INPUT:}  $RT^i, NT^i, l^i_n$, $u^i_{pn}$; }
    \STATE{{\bf [o]} verify $ u^i_{pn}$;}
        \STATE{$d^i_{pn} = ud^i_{pn}$;  ~~$ d_{min} = \infty; $}
        \FOR{{\bf each} $k \in N^i -\{ i \} $}       
            \IF{$(d^i_{pk} + l^i_k < d_{min})  \vee (d^i_{pk} + l^i_k = d_{min} \wedge |k| < |s_{new}|)$ }
                \STATE{$s_{new} = k; d_{min} = d^i_{pk} + l^i_k$; }
            \ENDIF
	\ENDFOR
	\IF{($d^i_{ps_{new}} < fd^i_p \vee [d^i_{ps_{new}}= fd^i_p \wedge |s_{new}| < |i|]$)}
		\LIF{$s^i_p \neq s_{new}$}{$s^i_p = s_{new}$; $a^i_p = a^i_{ps_{new}}$} 
		\IF{$d^i_p \neq d_{min}$}
			\STATE{$d^i_p = d_{min}; fd^i_p = min\{fd^i_p, d^i_p\}; V^i_p = \phi$;}
			\FOR{{\bf each} $k \in N^i -\{ i \} $} 
				\STATE{$uf^i_{pk} = 1$; $tf^i_{pk} =${\it UPDATE};}   
				\IF{($d^i_{kp} < fd^i_p \vee [d^i_{pk}= fd^i_p \wedge |k| < |i|]$)}
					\STATE{$V^i_p =  V^i_p \cup k$;}
				\ENDIF
			\ENDFOR   
			\LIF{$ut^i_{pn} =$ {\it QUERY}}{$tf^i_{pn} =${\it REPLY};}
		\ENDIF
	\ELSE
		\STATE{$mf^i_p = $ {\it ACTIVE;} $d^i_p = d^i_{ps^i_p} + l^i_{s^i_p};$}
		\LIFELS{($n = s^i_p \wedge ut^i_{pn} =$ {\it QUERY})}{$o^i_p = 3;$}{$o^i_p = 1;$}
		\LFOR{{\bf each} $k \in N^i -\{ i \} $} 
				{$uf^i_{pn} = 1$; $tf^i_{pn} =${\it QUERY};}    
	\ENDIF
\end{algorithmic}}       
\end{algorithm}
\end{figure}

 \vspace{-0.1in}	
When a router is in {\it ACTIVE} mode, it cannot change its successor or $fd^i_p$ until it receives  the replies to its {\it QUERY} from all its neighbors. After receiving all replies (i.e. $rf^i_{pn} = 0| \forall n \in N^i$), router $i$ becomes {\it PASSIVE} by resetting its feasible distance. The router then selects the new successor and sends {\it UPDATE} messages to its neighbors. More specifically, router $i$ sets $fd^i_p = \infty$ which insures that the router can find a new successor that satisfies {\it SRC} and then sets $fd^i_p = d^i_p = min\{d^i_{pn}+l^i_n | n \in N^i\}$ and becomes {\it PASSIVE}. 

 \vspace{-0.15in}
\begin{figure}[h]
\begin{algorithm}[H]
\caption{Processing routing messages in {\it ACTIVE} mode}
\label{alg:o0}
{\fontsize{9}{9}\selectfont
\begin{algorithmic}
    \STATE{\textbf{INPUT:}  $RT^i, NT^i, u^i_{pn}$; }
    \STATE{{\bf [o]} verify $ u^i_{pn}$;}
    \STATE{$d^i_{pn} = ud^i_{pn};$}

    \IF{$ut^i_{pn} =$ {\it REPLY}}
    	\STATE{$rf^i_{pn} = 0; lastReply = true;$}
	\FOR{{\bf each} $k \in N^i -\{ i \} $} 
		\LIF{$rf^i_{pk} = 0$}{$lastReply = false$;}
	\ENDFOR	
	\IF { $lastReply = true$}
		\LIF{$o^i_p = 1 \vee o^i_p = 3$} {$fd^i_p = \infty$}
		\STATE{Execute Algorithm \ref{alg:updateRTinf}}
       	\ENDIF
    \ELSIF{$ut^i_{pn} =$ {\it QUERY}}
	\IF{($o^i_p = 1 \vee o^i_p = 2$)} 
	    	\LIFELS{$n \neq s^i_p$}{$uf^i_{pn} = 1$; $tf^i_{pn} =${\it REPLY;}}{$o^i_p = 4;$}
	\ENDIF	
	\LIF{($o^i_p = 3 \vee o^i_p = 4$)}{$uf^i_{pn} = 1$; $tf^i_{pn} =${\it REPLY;}}
    \ENDIF
\end{algorithmic}}       
\end{algorithm}
\end{figure}

\vspace{-0.32in}
\begin{figure}[h]
\begin{algorithm}[H]
\caption{Update $RT^i_p$ }
\label{alg:updateRTinf}
{\fontsize{9}{9}\selectfont
\begin{algorithmic}
    \STATE{\textbf{INPUT:}  $RT^i, NT^i, l^i_n$, $u^i_{pn}$; }
                \STATE{$ d_{min} = \infty; $}
                \FOR{{\bf each} $k \in N^i -\{ i \} $}       
                \IF{$(d^i_{pk} + l^i_k < d_{min})  \vee (d^i_{pk} + l^i_k = d_{min} \wedge |k| < |s_{new}|)$ }
                        \STATE{$s_{new} = k; d_{min} = d^i_{pk} + l^i_k$; }
                    \ENDIF
            	\ENDFOR
            	\IF{($d^i_{ps_{new}} < fd^i_p \vee [d^i_{ps_{new}}= fd^i_p \wedge |s_{new}| < |i|]$)}
			\STATE{$o^i_p = 0; mf^i_p = PASSIVE$;}   
            		\LIF{$s^i_p \neq s_{new}$}{$s^i_p = s_{new}$;} 
            		\IF{$d^i_p \neq d_{min}$}
            			\STATE{$d^i_p = d_{min}; fd^i_p = min\{fd^i_p, d^i_p\}; V^i_p = \phi$;}
            			\FOR{{\bf each} $k \in N^i -\{ i \} $} 
            				\STATE{$uf^i_{pk} = 1$; $tf^i_{pk} =${\it UPDATE};}   
            				\IF{($d^i_{pk} < fd^i_p \vee [d^i_{pk}= fd^i_p \wedge |k| < |i|]$)}
            					\STATE{$V^i_p =  V^i_p \cup k$;}
            				\ENDIF
            			\ENDFOR
            			\LIF{$qf^i_{ps^i_p}(old) = 1$}{$tf^i_{pn} =${\it REPLY;}}
            		\ENDIF
            	\ELSE
            		\LIFELS{$o^i_p = 2$}{$o^i_p = 1$}{$o^i_p = 3;$}
            		\LFOR{{\bf each} $k \in N^i -\{ i \} $} 
            				{$uf^i_{pn} = 1$; $tf^i_{pn} =${\it QUERY};}    
            	\ENDIF
\end{algorithmic}}       
\end{algorithm}
\end{figure}

	If router $i$ receives a {\it QUERY} from a neighbor other than its successor while it is {\it ACTIVE}, it simply replies to its neighbor  with a {\it REPLY} message stating the current distance to the destination. The case of a router receiving a {\it QUERY} from its successor while it is {\it ACTIVE} is described subsequently in the context of multiple diffusing computations. {\it UPDATE} messages are processed and neighbor tables are updated, but the successor or distance is not changed until the router receives all the replies it needs to transition to the   {\it PASSIVE} mode. While a router is in {\it ACTIVE} mode, neither a {\it QUERY} nor an {\it UPDATE} can be sent.

\begin{figure*}[th]
\centering
\includegraphics[width=.99\textwidth]{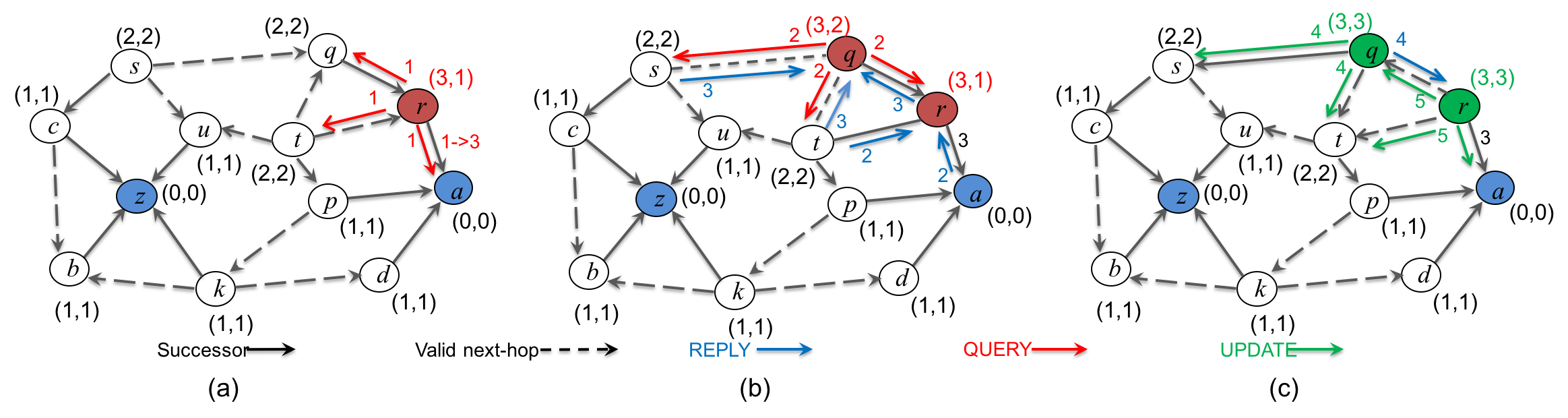}
\vspace{-0.1in}
\caption{DNRP Operation Example}
\label{fig:example}
\end{figure*}

	{\bf Handling Multiple Diffusing Computations:} 
Given that a router executes each local computation to completion, it 
handles multiple local computations for the same prefix one at a time. Similarly, 
a router handles multiple diffusing computation for the same  prefix by processing one computation at a time. An {\it ACTIVE} router $i$ can be in one of the following four states:  (1)  router $i$ originated a diffusing computation ($o^i_p = 1$), (2)  metric increase detected during {\it ACTIVE} mode ($o^i_p = 2$), (3) diffusing computation is relayed ($o^i_p = 3$), or (4) successor metric changed during {\it ACTIVE} mode ($o^i_p = 4$). If the router is in {\it PASSIVE} mode then its state is 0 (i.e.,  $o^i_p = 0$).
			
		Consider the case that  a router $i$  is {\it ACTIVE} and  in State 1 ($o^i_p = 1$). If the router  receives the last {\it REPLY} to its query, then it resets its feasible distance to infinity, checks  {\it SRC} to find the new successor, and sends an {\it UPDATE} to all its neighbors. On the other hand, if router $i$ detects a change in the link to its successor then it updates its neighbor table and sets $o^i_p = 2$.

	If  router $i$ is in State 2, receives the last {\it REPLY}, and can find a feasible successor using  {\it SRC} with the current feasible distance, then it becomes {\it PASSIVE} and sends an {\it UPDATE} to all its neighbors($o^i_p = 0$). Otherwise, it sends a {\it QUERY} with the current distance and sets $o^i_p = 1$. 

	Router $i$ uses the pending query flag ($qf^i_{pn}$) to keep track of the
replies that have been received for its {\it QUERY} regarding prefix $p$.
If router $i$ is in either  State 1 or 2 and  receives a {\it QUERY} from its current successor to the prefix, then  it sets $qf^i_{ps^i_p = 1}$ and transitions to State 4 (i.e., it sets $o^i_p = 4$).

			If a router  in {\it PASSIVE} mode receives a {\it QUERY} from its successor,
it searches for a new successor that satisfies  {\it SRC}. If it cannot find such a successor then it keeps the current successor, updates its distance,  and becomes {\it ACTIVE}. Then, the router sends {\it QUERY} to all of its neighbors and sets $o^i_p = 3$. 

	When router $i$ in state 3 receives {\it REPLY} from all of its neighbors, it resets its feasible distance, $fd^i_p = \infty$, selects a new successor, updates the $V^i_p$ and sends {\it UPDATE} to its neighbors and {\it REPLY} to its the previous successor. If the router  detects a link failure or a cost increase in the link to its current successor, the router sets $o^i_p = 4$ to indicate that a topology change occurred while the router is in {\it ACTIVE} mode.  A  router  handles the case  of the failure of  the link with its successor as if it had received a {\it REPLY} from its successor with $d^i_{ps^i_p} = \infty$.
	
	If router $i$ is in State 4, ($o^i_p = 4$) and it receives replies from all its neighbors, then it tries to find a feasible successor that satisfies  {\it SRC} with the current value of $fd^i_p$. If such a successor exists, the router  updates its successor, distance, and next hops for prefix $p$, and sends an {\it UPDATE} message to its neighbors as well as {\it REPLY} to the previous successor. Otherwise, it sets  $o^i_p = 3$ and sends a {\it QUERY} with the new distance.

	 While  router $i$   is in {\it ACTIVE} mode regarding a prefix, if a {\it QUERY} is received for the prefix  from a neighbor other than the current successor, the router updates the neighbor table and sends a {\it REPLY} to that neighbor. If a router  in {\it PASSIVE} mode receives a {\it QUERY}  from a neighbor other than the current successor, the router updates its neighbor table. If the feasibility condition is not satisfied anymore, the router sends a {\it REPLY} to the neighbor 
that provides  the current value $d^i_p$ before it starts its own computation. 

\subsection{Example of DNRP Operation}
	 
	 Figure \ref{fig:example} illustrates the operation of DNRP with a simple example.  The figure shows the routing information used for a single prefix when routers $a$ and $z$ advertise  that prefix and each link has unit cost. The tuple next to each router states the $distance$ and the $feasible~distance$ of the router for that prefix. The red, blue, and green arrows represent the  {\it QUERY},  {\it REPLY}, and  {\it UPDATE} messages respectively and the number next to the arrow shows the time sequence in which that message is sent.  Figure  \ref{fig:example} (a) shows the change in the cost of link $(r,a)$.  Router $r$ detects this change and becomes {\it ACTIVE} and sends {\it QUERY} to its neighbors. 

	   Router $q$ receives the {\it QUERY} from its successor and cannot find a feasible successor (Figure \ref {fig:example}(b)). Therefore, it becomes {\it ACTIVE} and sends a {\it QUERY} to its neighbors.   Router  $r$ receives {\it REPLY} from $a$ and $t$, and  a {\it QUERY} from $q$. Given that  $q$ is not a  successor for router $r$, $r$ sends {\it REPLY} to $q$. After receiving {\it REPLY} from routers $r$, $s$ and $t$, router $q$ becomes {\it PASSIVE} again and sends its {\it REPLY} to its previous successor, $r$. In turn, this means that  $r$ receives all the replies it needs, becomes {\it PASSIVE},  and resets its feasible distance. The operation of DNRP is such that only a portion of the routers are affected by the topology change.

\subsection{Routing to all instances of a prefix}

DNRP enables routers to maintain multiple loop-free routes to the nearest anchor of a name prefix. In some ICN architectures, such as NDN and CCNx, an anchor of a name-prefix may have some but not necessarily all the content corresponding to a given prefix. Therefore, simply routing to nearest replica may cause some data to be unreachable, and the ability to contact all anchors of a prefix is needed. To address this case, a  multi-instantiated destination spanning tree ({\it MIDST}) can be used alongside DNRP to support routing to all anchors of the same prefix. A MIDST is established in a distributed manner. Routers that are aware of multiple anchors for the same prefix exchange routing updates to establish the spanning tree between all anchors of a prefix. Once the {\it MIDST} is formed for a given prefix, the first router in the MIDST that receives a packet forwards it over the {\it MIDST} to all of the anchors.  The details  of how a  {\it MIDST} can be established  in DNRP are omitted for brevity; however, the approach is very much the same as that described in  \cite{JJMIDST2014}.
	 
	
\section{Correctness of DNRP}
\label{sec-correct}
The following theorems prove that DNRP is loop-free at every instant and considers each computation individually and in the proper sequence. From these results, the proof that  DNRP converges to shortest paths to prefixes is similar to the proof presented in \cite{jjdistribute} and due to space limitation is omitted.
We assume that each router receives and processes all routing messages correctly. This implies that each router processes messages from each of its neighbors in the correct order. 


 \vspace{0.075in}
\begin{theorem}
No routing-table loops can form in a network in which routers use {\it NSC} to select their next hops to prefixes.
\label{theorem:nsccorrectness}
\end{theorem}
\begin{IEEEproof}
	Assume for the sake of contradiction that  all routing tables are loop-free before time $t_l$ but  a routing-table loop  is formed for prefix $p$ at time $t_l$ when router $q$ adds its neighbor $n_1$ to its valid next-hop set $V^q_p$. Because the successor is also a valid next hop, router $q$ must either choose a new successor or add a new neighbor other than its current successor to its valid next-hop set at time $t_l$. We must show that 
the existence of a routing-table  loop is a contradiction  in either case. 

Let $L_p$ be the routing-table loop  consisting of $h$ hops starting at router $q$, ($L_p =\{q=n_{0,new}, n_{1,new},  n_{2,new}, \dots, $ $n_{h,new}\}$) where $n_{h,new} = q$, $n_{i+1,new} \in V^{n_i}_p$ for $0 \leq i \leq h$.

The time router $n_i$ updates its valid next-hop set to include $n_{i+1, new}$ is denoted by $t^{i}_{new}$. 
Assume that the last time router $n_{i}$ sent an {\it UPDATE} 
that was processed by 
its neighbor $n_{i-1}$, is  $t^{i}_{old}$. Router $n_i$ revisits valid next hops after any changes in its successor, distance, or feasible distance; therefore, 
$t^{i}_{old} \leq t^{i}_{new} \leq t_l$
and $d^{n_i}_{pn_{i+1}}(t_l) = d^{n_i}_{pn_{i+1}}(t_{old})$.
Also, by definition, at any time $t_i$, $fd^i_p(t_i) \leq d^i_p(t_i)$, and $fd^i_p(t_2) \leq fd^i_p(t_1)$ if $t_1 < t_2$. Therefore,
\begin{equation}
fd^i_p(t_2) \leq d^i_p(t_1) \mbox{~such that ~} t_1 < t_2 
\label{eq:fact1}
\end{equation}
If router $n_i$ selects a new successor at time $t^{i}_{new}$ then: 
\begin{equation}
d^{n_{i-1}}_{pn_{i}}(t_l) = d^{n_i}_{p}(t_{old}) 
\geq fd^{n_i}_{p}(t_{old}) 
\geq fd^{n_i}_{p}(t_{new})
\label{eq:cr1}
\end{equation}
Using {\it NSC} ensures  that
\begin{equation}
\begin{split}
(fd^{n_i}_{p}(t_{new}) > d^{n_{i}}_{pn_{i+1}}(t_l)) \\
\vee (fd^{n_i}_{p}(t_{new}) = d^{n_{i}}_{pn_{i+1}}(t_l) \wedge |n_i| > |n_{i+1}|)
\end{split}
\label{eq:cr2}
\end{equation}
From Eqs. (\ref{eq:cr1}) and  (\ref{eq:cr2}) we have:
\begin{equation}
\begin{split}
 ( d^{n_{i-1}}_{pn_{i}}(t_l) > d^{n_{i}}_{pn_{i+1}}(t_l) )\\
\vee ( d^{n_{i-1}}_{pn_{i}}(t_l) = d^{n_{i}}_{pn_{i+1}}(t_l) \wedge |n_{i}| > |n_{i+1} | )
\end{split}
\end{equation}
Therefore,  for  $0 \leq k \leq h$  in $L_p$ it is true that:
\begin{equation}
\begin{split}
( d^{n_{0}}_{pn_{1}}(t_l) > d^{n_{k}}_{pn_{k+1}}(t_l) )\\
\vee  ( d^{n_{0}}_{pn_{1}}(t_l) = d^{n_{k}}_{pn_{k+1}}(t_l) \wedge |n_{0}| > |n_{k} | )
\end{split}
\label{eq:cr3}
\end{equation}

 If $d^{n_{i-1}}_{pn_{i}}(t_l) > d^{n_{i}}_{pn_{i+1}}(t_l)$ in at least one hop in $L_p$   then it must be true that, for any given $k \in \{1,2, \dots,h\}$,  $d^{n_{k}}_{pn_{k+1}}(t_l) > d^{n_{k}}_{pn_{k+1}}(t_l)$, which is a contradiction. If at any hop in the $L_p$ it 
is true that  $d^{n_{i-1}}_{pn_{i}}(t_l) = d^{n_{i}}_{pn_{i+1}}(t_l)$, then $|k| > |k|$, which is also a contradiction. Therefore, no routing-table loop can be formed when routers use {\it NSC} to select their next hops to prefix $p$. 
\end{IEEEproof}

 \vspace{0.05in}
\begin{lemma}
 	A router that is not the origin of a diffusing computation sends a {\it REPLY} to its successor when it becomes {\it PASSIVE}.
\end{lemma}

\begin{IEEEproof}
 	A router that runs DNRP can be in either {\it PASSIVE} or {\it ACTIVE} mode for a prefix $p$ when it receives a {\it QUERY} from its successor regarding the prefix.
Assume that router $i$ is in {\it PASSIVE} mode when it receives a {\it QUERY} from its successor. If router $i$ finds a neighbor that satisfies {\it SRC}, then it sets its new successor and sends a {\it REPLY} to its old successor. Otherwise, it becomes {\it ACTIVE}, sets $o^i_p = 3$,  and sends a {\it QUERY} to all its neighbors.
Router $i$  cannot  receive a subsequent {\it QUERY} from its successor regarding the same prefix, until it sends a {\it REPLY} back to its successor. If the distance does not increase while router $i$  is {\it ACTIVE} then $o^i_p$ remains the same (i.e. $o^i_p = 3$). Otherwise,  router $i$  must set $o^i_p = 4$. In both cases router $i$ must send a {\it REPLY} when it becomes {\it PASSIVE}.

Assume that router $i$ is in {\it ACTIVE} mode when it receives a {\it QUERY} from its successor $s$.  Router $s$ cannot send another {\it QUERY} until it receives a {\it REPLY} from all its neighbors to its query, including router $i$. Hence, router $i$ must be the origin of the diffusing computation for which it is {\it ACTIVE}  when it receives the {\it QUERY} from $s$, which means that  $o^i_p = 1$ or $o^i_p = 2$. In both cases router $i$ sets $o^i_p = 4$ when it receives a {\it QUERY} form its successor $s$ and $s$  must send a {\it REPLY} in response to the  {\it QUERY} from $i$ because, $i$ is not the successor for $s$. 
After receiving the last {\it REPLY} from its neighbors, either router $i$ finds a feasible successor and sends a {\it REPLY} to $s$ ($o^i_p = 0$) or it propagates the diffusing computation forwarded by $s$ by sending a {\it QUERY} to its neighbors and 
setting  $o^i_p = 3$. Router $i$ then must send a {\it REPLY} to $s$ when it receives the last {\it REPLY}  for the  {\it QUERY} it forwarded from $s$.
	
	Hence, independently of its current mode,  router $i$ must send a {\it REPLY} to 
a  {\it QUERY} it receives from its successor when it becomes {\it PASSIVE}.
\end{IEEEproof}

\begin{lemma}
 	Consider a network that  is loop free before an arbitrary time $t$ and in which  a single diffusing  computation takes place.  If node $n_{i}$ is {\it PASSIVE} for prefix $p$ at that time, then  it must be true that
$	( d^{n_{i-1}}_{pn_{i}}(t) > d^{n_{i}}_{pn_{i+1}}(t) ) \vee 
	( d^{n_{i-1}}_{pn_{i}}(t) = d^{n_{i}}_{pn_{i+1}}(t) \wedge |n_i| > |n_{i+1} | )$ 
independently of the state of other routers in the 
chain of valid next hops  \{$n_{i-1}, n_i, n_{i+1}$\} for prefix $p$.

\label{lemma:uneq}
\end{lemma}
\begin{IEEEproof}
	Assume that router $n_i$ is {\it PASSIVE} and selects  router $n_{i+1}$ as a valid next hop. According to {\it NSC} it must be true that: 
\begin{equation}
\begin{split}
	 (d^{n_i}_{pn_{i+1}}(t)  <  fd^{n_i}_{p}(t) \leq d^{n_i}_{p}(t) ) \vee  \\
	 (d^{n_i}_{pn_{i+1}}(t) =  fd^{n_i}_{p}(t) \leq d^{n_i}_{p}(t)  \wedge |n_{i+1}| < |n_i|)
\end{split}
\label{eq:nsc1}
\end{equation}
Assume that $n_i$ did not reset $fd^{n_i}_p$ the last time $t_{new} < t$ when $n_i$ became {\it PASSIVE} and selected its successor  $s_{new}$ and updated its distance $d^{n_i}_p (t_{new}) =  d^{n_i}_p (t) $. If router $n_{i-1}$ processed  the message that router $n_i$ sent after updating its distance, then: $d^{n_{i-1}}_{pn_{i}}(t) = d^{n_i}_{p}(t_{new})$. Substituting  this equation in \ref{eq:nsc1} renders the result of this lemma. 

	On the other hand, If router $n_{i-1}$ did not process the message that router $n_i$ sent after updating its distance and before $t$, then $d^{n_{i-1}}_{pn_{i}}(t) = d^{n_i}_{p}(t_{old})$. Based on the facts that router $n_i$ did not reset its feasible distance and  Eq. \ref{eq:fact1} holds for this case. Therefore:
\begin{equation}
	d^{n_{i-1}}_{pn_i}(t) = d^{n_{i-1}}_{pn_i}(t_{old}) > fd^{n_i}_{p}(t)
\label{eq:ly1}
\end{equation}

	Now consider the case that $n_i$ becomes {\it PASSIVE} at time $t_{new}$  and changes its successor from $s_{old}$ to $s_{new}$ by reseting its feasible distance. The case that $n_{i-1}$ processed  the message that router $n_i$ sent after becoming {\it PASSIVE} is the same as before. Assume that $n_{i-1}$ did not process the message that $n_i$ sent at time $t_{new}$. 
Furthermore, assume that  router $n_i$ becomes {\it ACTIVE} at time $t_{old}$, with  a distance $d^{n_i}_{p}(t_{old}) =  d^{n_i}_{ps_{old}}+l^{n_i}_{s_{old}}$. Router $n_i$ cannot change its successor or experience any increment in its distance through $s_{old}$; hence, $d^{n_i}_{p}(t_{new})  \leq d^{n_i}_{p}(t_{old})$. On the other hand, the distance through the new successor must be the shortest and so $d^{n_i}_{p}(t_{new}) =  d^{n_i}_{ps_{new}}+l^{n_i}_{s_{new}} \leq d^{n_i}_{p}(t_{old})$.  Router $n_i$ becomes {\it PASSIVE} if it receives a {\it REPLY} from each of its neighbors including $n_{i-1}$. Therefore, $n_{i-1}$ must be notified about $d^{n_i}_{p}(t_{old})$ .   Therefore:
\begin{equation}
\begin{split}
	d^{n_{i-1}}_{pn_{i}}(t) = d^{n_i}_{p}(t_{old}) \geq d^{n_i}_p (t_{new}) =  d^{n_i}_p (t).
\end{split}
\end{equation}
Substituting this equation in \ref{eq:nsc1} renders the result of this lemma. Therefore, the lemma is true in all cases.
\end{IEEEproof}

\begin{lemma}
 	Consider a network that  is loop free before an arbitrary time $t$ and in which a single diffusing computation takes place. Let two network nodes $n_i$ and $n_{i+1}$ be such that $n_{i+1} \in V^{n_i}_p$. 
Independently of the state of these two nodes, it must be true that:
\begin{equation}
\begin{split}
	( fd^{n_{i}}_{p}(t) > fd^{n_{i+1}}_{p}(t) ) \vee \\
	( fd^{n_{i}}_{p}(t) = fd^{n_{i+1}}_{p}(t) \wedge |n_{i}| > |n_{i+1} | ) 
\end{split}
\label{eq:l2}
\end{equation}
\label{lemma:uneq2}
\end{lemma}

\begin{IEEEproof}
	Consider the case that router $n_i$ is {\it PASSIVE}, then from Lemma \ref{lemma:uneq} and the fact that routers select their next hops based on {\it NSC}, it must be true that:
\begin{equation}
\begin{split}
	( fd^{n_{i}}_p > d^{n_{i}}_{pn_{i+1}}(t) ) \vee \\
	(  fd^{n_{i}}_p = d^{n_{i}}_{pn_{i+1}}(t) \wedge |n_{i}| > |n_{i+1} | ) 
\end{split}
\label{eq:l1}
\end{equation}
	Consider the case that router $n_{i+1}$ is {\it ACTIVE}. Router $n_{i+1}$
cannot change its successor or increase its feasible distance. If router $n_i$ processed the last message that router $n_{i+1}$ sent before time $t$, then: $d^{n_{i}}_{pn_{i+1}}(t) = fd^{n_{i+1}}_p(t)$ and the lemma is true. 
Assume router $n_i$ did not process the last message that router $n_{i+1}$ sent before time $t$. Router $n_i$ must send a {\it REPLY} to  $n_{i+1}$ the last time that router $n_{i+1}$ became {\it PASSIVE} at time $t_p$ reporting  
a distance $d^{n_{i+1}}_{p}(t_{old}) =  d^{n_{i+1}}_{ps_{old}}+l^{n_{i+1}}_{s_{old}}$. 

If router $n_{i+1}$ did not reset its feasible distance since the last time it became passive, $fd^{n_{i+1}}$, then,  $d^{n_{i+1}}_{p}(t_{old}) \geq fd^{n_{i+1}}_{p}(t)$.
Consider the case that router $n_{i+1}$ resets  $fd^{n_{i+1}}$ the last time before $t$ that it becomes {\it PASSIVE}. Router $n_{i+1}$ cannot change its successor or experience any increment in its distance through its old successor, $s_{old}$. Hence, $d^{n_{i+1}}_{p}(t_{new})  \leq d^{n_{i+1}}_{p}(t_{old})$. On the other hand, the distance through the new successor must be the smallest among all neighbors including the old successor and so $d^{n_{i+1}}_{p}(t_{new}) =  (d^{n_{i+1}}_{ps_{new}}+l^{n_{i+1}}_{s_{new}}) \leq d^{n_{i+1}}_{p}(t_{old})$.  Router $n_{i+1}$ becomes {\it PASSIVE} if it receives a {\it REPLY} from each of its neighbors, including $n_{i}$. Therefore, $n_{i}$ must be notified about $d^{n_{i+1}}_{p}(t_{old})$ .   Therefore,
\begin{equation}
\begin{split}
	d^{n_{i}}_{pn_{i+1}}(t) = d^{n_{i+1}}_{p}(t_{old}) \geq d^{n_{i+1}}_p (t_{new}) \geq fd^{n_{i+1}}_p (t_{new})
\end{split}
\label{eq:ne1}
\end{equation}
The feasible distance $fd^{n_{i+1}}_p (t_{new})$ with $t_{new} < t$ cannot increase until router $n_{i+1}$ becomes {\it PASSIVE} again; therefore,$fd^{n_{i+1}}_p (t_{new}) \geq fd^{n_i}_p (t)$. 
The result of the lemma follows in this case by substituting this  result in Eqs. (\ref{eq:ne1}) and Eq. (\ref{eq:l1}).  

	Now consider the case that router $n_{i+1}$ is {\it PASSIVE}.  If router $n_i$ processed the last message that router $n^{i+1}$ sent before time $t$, then $d^{n_{i}}_{pn_{i+1}}(t) = d^{n_{i+1}}_p(t) \geq fd^{n_{i+1}}_p(t)$ and the lemma is true. Now consider the case that  router $n_i$ did not process the last message router $n^{i+1}$  sent  before time $t$. If router $n_{i+1}$ did not reset $fd^{n_{i+1}}$ then  $d^{n_{i+1}}_{p}(t_{old}) \geq fd^{n_{i+1}}_{p}(t)$. On the other hand, if router $n_{i+1}$ resets  $fd^{n_{i+1}}$ then we can conclude that 
 $fd^{n_{i+1}}_p (t_{new}) \geq fd^{n_i}_p (t)$ and $|n_i| > |n_{i+1}|$ using an argument similar to one we used for the {\it ACTIVE} mode. Hence, the lemma is true for all  cases.
\end{IEEEproof}

 {\it NSC} and {\it SRC} guarantees loop-freedom at every  time instant.  If we consider the link form router $i$ to its valid next hop with respect to a specific prefix as a directed edge, then the graph containing all this directed links is a directed acyclic graph (DAG) with respect to that specific prefix. The DAG representing the relationship of valid next hops regarding prefix $p$ is denoted by $D_p$.
	
\begin{figure*}[th]
\centering
\includegraphics[width=.95\textwidth]{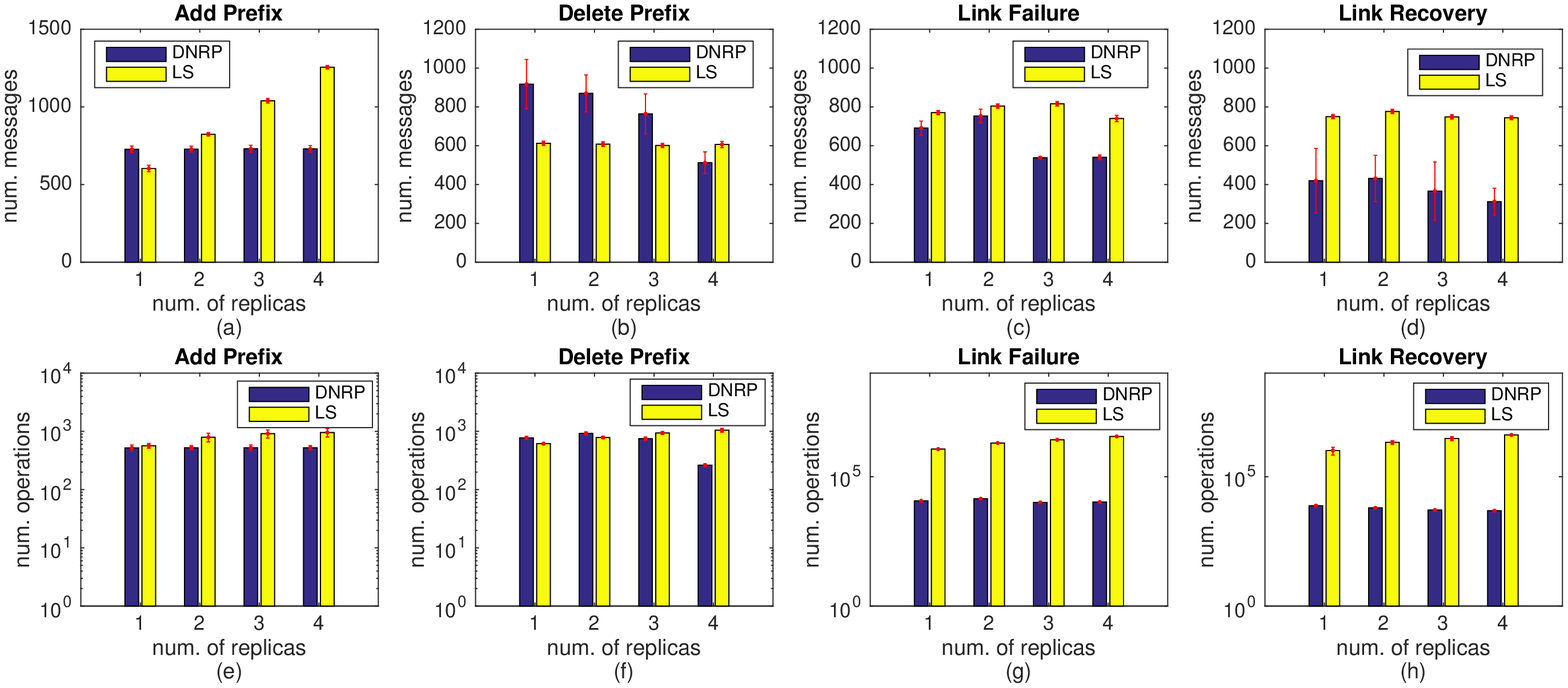}
\vspace{-0.1in}
\caption{Simulation results showing average number of messages and average number of operations vs number of replicas}
\label{fig:sim}
\end{figure*}

\begin{lemma}
 	If routers are involved in a single diffusing computation then $D_p$ is loop-free at every instant. 
\end{lemma}
\begin{IEEEproof}
	Assume for the sake of contradiction  that $D_p$ is loop-free before an arbitrary time $t$ and a loop $L_p$ consisting of $h$ hops is created at time $t_l > t$ when router $q$ updates  $V^q_p$ after processing an input event. Assume that $L_p = \{n_1, n_2, \dots, n_h$ \} is  the loop created, where $n_{i+1} \in V^{n_i}_p$ for $1 \leq i \leq h$ and $n_1 \in V^{n_h}_p$. If router $n_1$ changes its next hop as a result of changing its  successor, it must be in  {\it PASSIVE} mode at time $t_l$ because an {\it ACTIVE} router cannot change its successor or update its next-hop set. 
	
	If all routers in $L_p$ are {\it PASSIVE} at time $t_l$, either all of them have always been {\it PASSIVE} at every instant  before $t_l$, or at least one of them was {\it ACTIVE} for a while and became  {\it PASSIVE} before $t_l$. If no router was ever {\it ACTIVE} before time $t_l$, it follows from Theorem \ref{theorem:nsccorrectness} that updating $V^n_p$ cannot create loop. Therefore, for router $n_1$ to create a loop, at least one of the routers must have been {\it ACTIVE} before time $t$.
	
	If all routers are in {\it PASSIVE} mode at time $t$, traversing $L_p$ and applying Theorem \ref{eq:l1} leads to the erroneous conclusion that  either $d^{n_1}_p > d^{n_1}_p$ or $|{n_1}| > |{n_1}|$. Therefore updating $V^{n_1}_p$ cannot create a loop if all routers in the $L_p$ are {\it PASSIVE} at time $t$.
	
	Assume that only one diffusing computation is taking place at time $t_l$. Based on Lemma \ref{lemma:uneq2} 
traversing  loop $L_p$ 
leads to the  conclusion that  either $fd^{n_i} > fd^{n_{i}}$ or $|n_i| > |n_i|$, which is  a contradiction.  Therefore, if only a single diffusing computation takes place, then $L_p$ cannot be formed when routers use {\it SRC} and {\it NSC} along with difusing computations to select next hops to reach the destination prefix.
\end{IEEEproof}

At steady state, the graph containing the successors and connected links between them, must create a tree. The tree containing successors that are {\it ACTIVE} regarding prefix $p$ and participating in a diffusion computation started form router $i$ at time $t$ are called diffusing tree ($T_{pi}(t)$). 
	    
\begin{theorem}
	DNRP considers each computation individually and in the proper sequence.
\end{theorem}

\begin{IEEEproof}
	Assume router $i$ is the only router that has started a diffusing computation up to time $t$. If router $i$ generates a single diffusion computation, the proof is immediate. Consider the case that router $i$ generates multiple diffusing computations. Any router that is already participating in the current diffusing computation (routers in the $T_{pi}$, including the router $i$) cannot send a new {\it QUERY} until it receives all the replies to the {\it QUERY} of  the current computation and becomes {\it PASSIVE}. Note that each router processes each event in order. Also, when a router becomes {\it PASSIVE}, it must send a {\it REPLY} to its successor, if it has any. Therefore, all the routers in $T_{pi}$ must process each diffusing computation individually and in the proper sequence.
	
	Consider the case that multiple sources of diffusing computations exist regarding prefix $p$ in the network. Assume router $i$ is {\it ACTIVE} at time $t$. Then either router $i$ 
is the originator of the diffusing computation ($o^i_p = $ 1 or 2), or received a {\it QUERY} from its successor ($o^i_p = $ 3 or 4). If $o^i_p = $ 1 or 3, the router must become {\it PASSIVE} before it can send another {\it QUERY}. If the router is the originator of the computation ($o^i_p = $ 1 or 2) and receives a {\it QUERY} form its successor, it  holds that {\it QUERY} and sets $o^i_p = $4. Therefore, all the routers in the $T_{pi}$ remain in the same computation. Router $i$ can forward the new {\it QUERY} and become the part of the larger $T_ps$ only after it 
receives a {\it REPLY} form each of  its neighbors for the current diffusing computation. If router $a$ is {\it ACTIVE} and receives a {\it QUERY} from its neighbor $k \neq s^a_p$, then it sends a {\it REPLY} to its neighbor before creating a diffusing computation, which means that $T_{pa}$ is not part of the {\it ACTIVE} $T_p$ to which  $k$ belongs. Therefore, any two {\it ACTIVE} $T_{pi}$ and $T_{pj}$ have an empty intersection at any given time, it thus follows from the previous case that the Theorem is true.
\end{IEEEproof}

\section{Performance Comparison}
\label{sec-sim}
We compare DNRP with a link-state routing protocol given that NLSR  \cite{nlsr2} is  based on link states and is the routing protocol advocated in NDN, one of the leading ICN architectures. We implemented DNRP and an idealized version of NLSR, which we simply call ILS (for ideal link-state), in ns-3 using the needed extensions to support content-centric networking \cite{SCoNET}.
In the simulations,  ILS  propagates update messages using the intelligent flooding mechanism. There are two types of Link State Advertisements (LSA): An adjacency LSA  carries information regarding a router, its neighbors, and connected links; and a prefix LSA  advertises name prefixes, as specified in \cite {nlsr2}. 
For convenience, DNRP sends HELLO messages between neighbors  to detect changes in the sate of nodes and links. However, HELLO's can be omitted in a real implementation 
and  detecting node adjacencies can be done my monitoring packet forwarding success in the data plane.

The AT\&T topology \cite{Heckmann2003} is used because it is a realistic topology for simulations 
that  mimic part of the Internet topology. It has 154 nodes and 184 links.  A node has 2.4 neighbors on average. In the simulations, the cost of a link is set to one unit, and 30 nodes are selected as anchors that advertise 1200 unique name prefixes.  We generated test cases consisting of single link failure and recovery, and a single prefix addition and deletion. 

	
	To compare the computation and communication overhead of DNRP and ILS,  we measured the number of routing messages transmitted over the network and the number of operations executed by each routing protocol. The number of messages for ILS  includes the number of HELLO messages, Adjacency LSAs, and Prefix LSAs. For DNRP, this measurement indicates the total number of all the routing messages transmitted as a result of any changes. The operation count is incremented whenever an event occurs, and statements within a loop are executed. 

	The simulation results comparing DNRP with ILS are depicted in Figure \ref{fig:sim}. In each graph, the horizontal axes is the average number of anchors per prefix, i.e., the number of anchors that advertise the same prefix to the network.  We considered four scenarios: adding a new prefix to the network; deleting one prefix from one of the replicas; a single link failure; and a single link recovery. Hence, ILS incurs  the same signaling overhead independently of how many LSA's are carried in an update.
Figures (\ref{fig:sim}a - \ref{fig:sim}d) show the number of messages transmitted in the whole network while Figures (\ref{fig:sim}e - \ref{fig:sim}h) show the number of operations each protocol executed after the change. The number of operations in the figure is in logarithmic scale. 

ILS  advertises prefixes from each of the replicas to the whole network. As the number of replicas increases, the number of messages increases, because each replica advertises its own Prefix LSA. In DNRP,  adding a new prefix affects nodes in small regions and hence the number of messages and operations are fewer than in  ILS. Deleting a prefix from one of the replicas results in several diffusing computations in DNRP, which results in more
signaling. However, the number of messages decreases as the number of replicas increases, because the event affect fewer routers. In ILS one Prefix LSA will be advertised  for each deletion. The computation of prefix deletion is comparable; however, DNRP imposes less computation overhead when the number of replicas reach 4.

DNRP has less communication overhead compared to ILS  after a link recovery or a link failure. The need to execute  Dijkstra's shortest-path first  for each neighbor results in ILS requiring more computations than DNRP. 
DNRP outperforms NLSR for topology changes as well as adding  a new prefix. 

\section{Conclusion}
	
We introduced  the first name-based  content routing protocol based on diffusing computations (DNRP) and  proved that it  provides loop-free multi-path routes  to multi-homed name prefixes at every instant.  
Routers that run DNRP do not require to have knowledge about the network topology, use complete paths to content replicas,  know about all the sites storing replicas of named content, or use periodic updates. DNRP has better performance compared to link-state  routing protocols when topology changes occur or new prefixes are introduced to the network. 
A real implementation of DNRP would not require the  use of HELLO's used in our simulations, and hence its overhead is far less than routing protocols that rely on 
LSA's validated by sequence numbers, which require periodic updates to work correctly.

\section{Acknowledgments}
\vspace{-0.02in}
This work was supported in part by the Baskin Chair of Computer Engineering at UCSC.

\bibliographystyle{IEEEtran}



\begin{thebibliography}{88}\setlength{\itemsep}{.5ex}
\vspace{-0.03in}


\bibitem{Bari2012}
M.~Bari et al., ``{A Survey of
  Naming and Routing in Information-Centric Networks},''   \emph{IEEE
  Communications Magazine}, vol.~50, no.~12, pp. 44--53, Dec. 2012. 

\bibitem{hlvr}
J. Behrens and J.J. Garcia-Luna-Aceves, ``Hierarchical Routing Using
Link Vectors,'' {\em Proc. IEEE INFOCOM '98}, March 1998.


\bibitem{Choi2011}
J.~Choi et al., ``{A Survey on Content-Oriented
  Networking for Efficient Content Delivery},'' \emph{IEEE Communications
  Magazine},  March 2011. 



\bibitem{jjdistribute}
J.~J. Garcia-Luna-Aceves, ``A Distributed, Loop-Free, Shortest-Path Routing Algorithm,'' 
  \emph{Proc. IEEE INFOCOM `88}, Mar 1988.


\bibitem{dual}
J.~J. Garcia-Luna-Aceves, ``Loop-Free Routing Using Diffusing Computations,''
  \emph{IEEE/ACM Transactions on Networking}, 1993.

\bibitem{icnp98}
J.J. Garcia-Luna-Aceves and M. Spohn, ``Scalable Link-State Internet
Routing,'' {\em Proc. ICNP `98}, Oct.  1998.

\bibitem{wild}
  J.J. Garcia-Luna-Aceves,   
``System and Method for Discovering Information Objects and Information Object Repositories in Computer Networks," U.S. Patent 8,572,214, October 29, 2013.


\bibitem{garcia2014name}
J.~J. Garcia-Luna-Aceves, ``Name-Based Content Routing in Information Centric  Networks Using Distance Information,'' in \emph{Proc. ACM ICN `14},   2014.


\bibitem{JJMIDST2014}
J.~J. Garcia-Luna-Aceves, ``Routing to Multi-Instantiated Destinations:
  Principles and Applications,'' \emph{Proc. IEEE ICNP 2014},  2014.


\bibitem{garcia2015fault}
J.~J. Garcia-Luna-Aceves, ``A Fault-Tolerant Forwarding Strategy for
  Interest-Based Information Centric Networks,''  \emph{Proc. IFIP Networking `15}, 2015.


\bibitem{Katsaros2012}
K.~V. Katsaros et al., ``On Inter-Domain Name Resolution for Information-Centric Networks," {\em Proc. Networking 2012},  May 2012.

\bibitem{Koponen2007}
T.~Koponen et al.,  ``A Data-Oriented (and Beyond) Network Architecture,'' {\em Proc. ACM SIGCOMM `07}, 2007.

\bibitem{nlsr2}
 V.~Lehman et al.,  ``A Secure Link State Routing Protocol for NDN,'' {\em IEEE Access}, Jan. 2018




\bibitem{mob}
Mobility first project. [Online]. Available:
  \url{http://mobilityfirst.winlab.rutgers.edu/}

\bibitem{PURSUIT}
Publish subscribe internet technology ({PURSUIT}) project. [Online]. Available:
  \url{http://www.fp7-pursuit.eu/PursuitWeb/}

\bibitem{gold} 
    J. Raju et al., 
 ``System and Method for
Information Object Routing in Computer Networks," U.S. Patent 7,552,233, June 23, 2009 
  
  \bibitem{nasr} 
M. Spohn and J.J. Garcia-Luna-Aceves, ``Neighborhood Aware Source
Routing,'' {\em Proc. ACM MobiHoc 2001}, Oct. 2001.

\bibitem{sail}
Scalable and adaptive internet solutions ({SAIL}) project. [Online]. Available:
  \url{http://www.sail-project.eu/}

\bibitem{SCoNET}
J.~Mathewson et al., 
  ``Sconet : Simulator content networking,''  \emph{CCNxCon}, 2015.

\bibitem{Heckmann2003}
O.~Heckmann et al., 
  ``On realistic network topologies for simulation" \emph{MoMeTools '03} , 2003.

\end{thebibliography}



 {\fontsize{6}{6}\selectfont

\end{document}